\documentclass[12pt, prd, showpacs]{revtex4}
\usepackage{amssymb}
\usepackage{amsmath}

\input{tcilatex}

\begin{document}

\title{Unbounded energies of debris from head-on particle collisions near
black holes }
\author{O. B. Zaslavskii}
\affiliation{Department of Physics and Technology, Kharkov V.N. Karazin National
University, 4 Svoboda Square, Kharkov, 61077, Ukraine}
\affiliation{Institute of Mathematics and Mechanics, Kazan Federal University, 18
Kremlyovskaya St., Kazan 420008, Russia}
\email{zaslav@ukr.net }

\begin{abstract}
If two particles move toward a black hole and collide near the horizon, the
energy $E_{c.m.}$ in the centre of mass can grow unbounded. This is a
so-called Ba\~{n}ados-Silk-West (BSW) effect. One of problems creating
obstacles to the possibility of its observation consists in that individual
energy $E$ of a fragment at infinity remains finite because of redshift. We
show that in the case of head-on collision, debris may have unbounded energy 
$E$. An essential ingredient of this scenario is a particle moving away from
a\ black hole in the near-horizon region. It can appear due to precedent
collision that implies multiple scattering.
\end{abstract}

\maketitle

\section{Introduction}

Several years ago, an interesting observation was made by Ba\~{n}ados, Silk
and West (the BSW effect, after names of its authors). If two particles
collide near a rotating black hole, the energy of collision in the centre of
mass $E_{c.m.}$ may grow unbounded \cite{ban}. This effects is important
from the theoretical point of view. It opens new channels of reactions
forbidden in the laboratory conditions. The BSW effect can also, in
principle, influence the geometry of space-time and, perhaps, the fate of \
black hole due to strong backreaction. It has also potential consequences in
astrophysics in what concerns physics of accretion disc, relativistic plasma
near black holes, etc. If the energy at infinity can be also made large,
this suggests the way to extract the energy from black holes due to the
collisional version of the Penrose process \cite{pen}.

However, there is difficulty that creates obstacle to direct registering
such an effect. Although in the point of collision the energy $E_{c.m.}$ can
be very large, the redshift compensates this effect from the viewpoint of an
remote observer, so the individual Killing energy $E$ of particles coming to
infinity turns out to be quite modest \cite{p} - \cite{z}. In particular,
for the Kerr metric, net outcome is less than 150 \% as compared to the
initial energy.

Meanwhile, there are some hints that the energy extraction efficiency can be
enhanced significantly in some situations. It was shown in \cite{rn} (and
confirmed by another methods in \cite{esc}) that for the Reissner-Nordstr%
\"{o}m black hole the upper limit on the energy of debris of collision does
not exist. There also some indications in the favour of unbounded $E$ for
geodesics at high inclinations to the equatorial plane of motion \cite{debr}%
. Significant enhancement of this energy for nonequatorial motion was found
in \cite{neq}.

Quite recently, one more work appeared \cite{card} in which it was claimed
that energies $E$ of debris may be not only very large but, in principle,
unbound. The results of this paper are presented in the form of figures
describing numerical calculations. The goal of this paper is to give
analytical approach to the problem that shows that unbound $E$ are indeed
possible. In doing so, the following crucial difference between the BSW
effect and the process under discussion comes into foreground. The key
moment in the BSW effect consists in that both particles move towards a
black hole. Then, this effect occurs if parameters of one of two particles
are fine-tuned. Meanwhile, if one particle moves away from the black hole
and the other one in the inward direction, so head-on collision occurs, $%
E_{c.m.}$ is always unbounded in the vicinity of the horizon \cite{pir1} - 
\cite{pir3}. The net outcome is finite in the standard Penrose process.
However, for the collisional one the situation may change. We show that in
the scenario when after head-on collisions particles move again in the
opposite directions, there are no upper bounds on $E$ at all. Thus, this
increases hopes to observe the result of collision in the laboratory. It is
worth noting that scenarios with high energy head-on collisions were
considered earlier but in the context where black holes were absent (in
particular, for naked singularities or star-like configurations) \cite{kerr}
- \cite{head}. Throughout the paper the fundamental constants $G=c=1$.

\section{Basic equations}

Let us consider the axially symmetric black hole metric 
\begin{equation}
ds^{2}=-N^{2}dt^{2}+g_{\phi \phi }(d\phi -\omega dt)^{2}+dn^{2}+g_{\theta
\theta }d\theta ^{2}\text{,}
\end{equation}%
where the metric coefficients do not depend on $t$ and $\phi $. Then, the
angular momentum $L=mu_{\phi }$ and the energy $E=-mu_{0}$ \ are conserved.
Here, $u^{\mu }$ is the four-velocity of a particle. The horizon lies at $%
N=0 $. For simplicity, let us consider motion within the equatorial plane $%
\theta =\frac{\pi }{2}$. Expressing $u^{\mu }$ in terms of $u_{\mu }$, using
the conservations law and the normalization conditions $u^{\mu }u_{\mu }=-1$%
, one obtains that 
\begin{equation}
mu^{0}=\frac{X}{N^{2}}\text{, }  \label{t}
\end{equation}%
\begin{equation}
mu^{3}=\frac{L}{g}+\frac{\omega X}{N^{2}}\text{,}
\end{equation}%
\begin{equation}
mu^{1}=\varepsilon \frac{Z}{N}\text{,}  \label{u1}
\end{equation}%
\begin{equation}
X=E-\omega L\text{, }Z=\sqrt{X^{2}-N^{2}(m^{2}+\frac{L^{2}}{g})},  \label{zu}
\end{equation}%
where $\varepsilon =+1$ for an outgoing particle and $\varepsilon =-1$ for
an ingoing one. The forward in time condition $u^{0}>0$ entails $X\geq 0$.

If two particles collide, we assume that the conservation laws are obeyed in
the point of collision, so

\begin{equation}
E_{1}+E_{2}=E_{3}+E_{4}\text{,}  \label{e}
\end{equation}%
\begin{equation}
L_{1}+L_{2}=L_{3}+L_{4},  \label{l12}
\end{equation}%
\begin{equation}
\varepsilon _{1}Z_{1}+\varepsilon _{2}Z_{2}=\varepsilon
_{3}Z_{3}+\varepsilon _{4}Z_{4},  \label{ez}
\end{equation}%
where Eq. (\ref{ez}) has the meaning of the conservation of the radial
momentum. It also follows from (\ref{e}) and (\ref{l12}) that%
\begin{equation}
X_{1}+X_{2}=X_{3}+X_{4}\text{.}  \label{x}
\end{equation}

It is worth stressing that individual energies $E_{i}$ are finite. It is the
energy in the centre of mass frame $E_{c.m.}$ which can, in principle,
diverge.

When two particles 1 and 2 collide, their energy in the centre of mass frame
can be defined according to%
\begin{equation}
E_{c.m.}^{2}=-P_{\mu }P^{\mu }\text{,}
\end{equation}%
$P^{\mu }=p_{1}^{\mu }+p_{2}^{\mu }$, is the total momentum, $p_{i}^{\mu
}=m_{i}u_{i}^{\mu }$, $m_{i}$ is the mass, index $i$ labels particles. Then,%
\begin{equation}
E_{c.m.}^{2}=m_{1}^{2}+m_{2}^{2}+2m_{1}m_{2}\gamma \text{,}
\end{equation}%
where the Lorentz factor of relative motion%
\begin{equation}
\gamma =-u_{1\mu }u^{2\mu }\text{.}
\end{equation}%
Then, using equations of motion (\ref{t}) - (\ref{u1}), one finds

\begin{equation}
\gamma =\frac{X_{1}X_{2}-\varepsilon _{1}\varepsilon _{2}Z_{1}Z_{2}}{%
m_{1}m_{2}N^{2}}-\frac{L_{1}L_{2}}{m_{1}m_{2}g_{H}}\text{.}  \label{ga}
\end{equation}

It is seen from (\ref{ga}) that for the motion in the opposite direction ($%
\varepsilon _{1}\varepsilon _{2}=-1$) $\gamma $ becomes unbounded if
collision occurs close to the horizon, where $N$ is small.

\section{Near-horizon expansions}

According to terminology accepted in works on high energy collisions near
black holes, we call particle usual if $X_{H}>0$ (critical, if $X_{H}=0),$%
where subscript "H" means that the corresponding quantity is calculated on
the horizon. We assume that all particles participating in collision are
usual, so in the near-horizon region we obtain

\begin{equation}
Z=X-\frac{1}{2X}(m^{2}+\frac{L^{2}}{g_{H}})N^{2}+O(N^{3})\text{.}  \label{uz}
\end{equation}%
In turn, the quantity $X$ can be expanded in the Taylor series. In this
region, the expansion of the coefficient $\omega $ near the extremal horizon
takes the general form 
\begin{equation}
\omega =\omega _{H}-B_{1}N+B_{2}N^{2}+O(N^{3})  \label{om}
\end{equation}%
where $\omega _{H}$ is the horizon value of $\omega $ and $B_{i}$ is some
model-dependent coefficient \cite{t}. Say, for the Kerr metric, expansion
goes in powers of $r-r_{+}$, where $r$ is the Boyer-Lindquist coordinate and 
$r_{+}$ is the horizon radius. For extremal black holes, $N\sim r-r_{+},$ so
expansion in terms of $r-r_{+}$ is equivalent to the expansion in terms of $%
N $. For the nonextremal black holes, $N^{2}\sim r-r_{+}$, so expansion
would start from the terms $N^{2}$ (see \cite{t} for details). In
particular, for the extremal Kerr metric, $B_{1}=M^{-1}$, $B_{2}=\frac{1}{2}%
M^{-1}$, where $M $ is the black hole mass \cite{circ}.

As a result,%
\begin{equation}
X=X_{H}+B_{1}LN-B_{2}LN^{2}+O(N^{3})\text{.}  \label{xn}
\end{equation}

The similar expansion is valid for the metric coefficient $g$:%
\begin{equation}
g=g_{H}+g_{1}N+g_{2}N^{2}+O(N^{3})\text{.}  \label{g}
\end{equation}

It follows from (\ref{uz}) and (\ref{ez}) that in the point of collision,
neglecting terms of the order $N^{2}$, we have

\begin{equation}
\varepsilon _{1}X_{1}+\varepsilon _{2}X_{2}=\varepsilon
_{3}X_{3}+\varepsilon _{4}X_{4}\text{.}  \label{x1}
\end{equation}

\section{Scenarios of collision}

Below, we assume that $\varepsilon _{1}=-1,\varepsilon _{2}=+1$. It means
that particle 1 moves towards a black hole where it meets particle 2 moving
away from a black hole. Particle 3 represents a fragment registered at
infinity, provided it escapes, so $\varepsilon _{3}=+1$. There are two
scenarios depending on $\varepsilon _{4}.$

\subsection{$\protect\varepsilon _{4}=-1$}

Then, it follows from (\ref{x1}) that, with a given accuracy, in the point
of collision

\begin{equation}
-X_{1}+X_{2}=X_{3}-X_{4}\text{.}
\end{equation}

Taking into account (\ref{x}), (\ref{xn}) we have 
\begin{equation}
(X_{1})_{H}=(X_{4})_{H}\text{,}  \label{14}
\end{equation}%
\begin{equation}
(X_{2})_{H}=(X_{3})_{H}\text{,}  \label{23}
\end{equation}

Equating in (\ref{ez}) the terms of the order $N^{2}$ and taking into
account (\ref{14}) and (\ref{23}), we get%
\begin{equation}
\frac{1}{2(X_{1})_{H}}(m_{1}^{2}+\frac{L_{1}^{2}}{g_{H}})-\frac{1}{%
2(X_{2})_{H}}(m_{2}^{2}+\frac{L_{2}^{2}}{g_{H}})=\frac{1}{2(X_{4})_{H}}%
(m_{4}^{2}+\frac{L_{4}^{2}}{g_{H}})-\frac{1}{2(X_{3})_{H}}(m_{3}^{2}+\frac{%
L_{3}^{2}}{g_{H}}).  \label{ml}
\end{equation}

We want to obtain unbounded energy $E_{3}>0$ for finite values of $E_{1}$, $%
E_{2}$, $L_{1}$, $L_{2}$ and finite masses. Because of (\ref{e}), this
entails that $E_{4}$ must be negative and large. Then, according to (\ref{zu}%
), the condition $X_{4}>0$ entails that $L_{4}$ is negative and large. So, $%
L_{3}$ is positive and big. Writing $L_{4}=L-L_{3}$, where $L$ is the total
angular momentum, we have from (\ref{ml}) that%
\begin{equation}
\frac{1}{2g_{H}}L_{3}^{2}(\frac{1}{X_{1}}-\frac{1}{X_{2}})_{H}-\frac{LL_{3}}{%
g_{H}(X_{1})_{H}}+\frac{m_{4}^{2}}{2(X_{1})_{H}}-\frac{m_{3}^{2}}{%
2(X_{2})_{H}}+\frac{1}{2(X_{2})_{H}}(m_{2}^{2}+\frac{L_{2}^{2}}{g_{H}})-%
\frac{1}{2(X_{1})_{H}}(m_{1}^{2}+\frac{L_{1}^{2}}{g_{H}})=0.
\end{equation}

In general, unbound $L_{3}$ are incompatible with the finiteness of all
other quantities. However, let 
\begin{equation}
L=0  \label{l0}
\end{equation}%
and, additionally, 
\begin{equation}
(X_{1})_{H}=(X_{2})_{H}.  \label{x12}
\end{equation}
Then, the potentially growing terms cancel, and we are left with a simple
condition%
\begin{equation}
m_{4}^{2}-m_{3}^{2}+m_{2}^{2}-m_{1}^{2}=0\text{.}
\end{equation}

For example, this can be satisfied for $m_{1}=m_{2}=m_{3}=m_{4}$.

\section{$\protect\varepsilon _{4}=+1$}

Then, instead, we obtain%
\begin{equation}
\frac{1}{2X_{1}}(m_{1}^{2}+\frac{L_{1}^{2}}{g_{H}})-\frac{1}{2X_{2}}%
(m_{2}^{2}+\frac{L_{2}^{2}}{g_{H}})=\frac{1}{2X_{4}}(m_{4}^{2}+\frac{%
L_{4}^{2}}{g_{H}})+\frac{1}{2X_{3}}(m_{3}^{2}+\frac{L_{3}^{2}}{g_{H}})
\end{equation}

As the right hand side is positive, in the limit \thinspace $%
L_{3}\rightarrow \infty $, it also diverges without compensation, whereas
the left hand side is finite, so this scenario is inappropriate for our
purposes. Moreover, the case when both particles 3 and 4 escape to infinity
is irrelevant for our purposes since both of them would have positive
energy, hence no energy extraction from a black hole.

\section{Nonequatorial motion}

The results can be generalized to nonequatorial motion in a straightforward
manner. Then, instead of (\ref{zu}), we have%
\begin{equation}
Z=X^{2}-N^{2}[m^{2}(1+g_{\theta \theta }\dot{\theta}^{2})+\frac{L^{2}}{g}]%
\text{,}  \label{z2}
\end{equation}%
where $\dot{\theta}=\frac{d\theta }{d\tau }$,~ $\tau $ is the proper time.

Assuming that $g_{\theta \theta }\dot{\theta}^{2}$ remains finite for all
particles, we see that this adds a finite correction to the large terms with 
$L^{2}$ in (\ref{z2}) and (\ref{ml}), so the main conclusions remain
unaffected.

\section{Summary and discussion}

We found that the energy measured at infinity can become unbound, provided
the conditions (\ref{l0}) and (\ref{x12}) are satisfied. This is a
counterpart of fine-tuning in the BSW effect \cite{ban}. More subtle point
consists in the question, how one can obtain a particle which near the black
hole horizon moves not towards a black hole but in the opposite direction.
The simplest and a natural kind of this a scenario can be realized when such
a a particle appears as a result of precedent collision. Thus although this
scenario cannot be realized in a straightforward manner, this is possible
due to multiple scattering. It is worth reminding that multiple scattering
scenario is necessary ingredient in another context - for the BSW effect
near nonextremal black holes \cite{gp}. Now, it can be used for extremal
black holes and head-on collisions. The present results can be considered as
analytical proof of numerical findings of \cite{card}. We also agree with 
\cite{card} in the role of multiple scattering.

In combination with \cite{rn} - \cite{card}, the present results show that
the question about scenarios with high energy debris from collisions near
black holes needs further treatment.

\begin{acknowledgments}
This work was funded by the subsidy allocated to Kazan Federal University
for the state assignment in the sphere of scientic activities.
\end{acknowledgments}

\end{document}